\documentclass[%
 reprint,
superscriptaddress,
preprintnumbers,
nofootinbib,
 amsmath,amssymb,
 aps,
 prl,
longbibliography,
twocolumn 
]{revtex4}  

\usepackage[pdftex]{graphicx}
\usepackage[utf8]{inputenc}
\usepackage{dcolumn}
\usepackage{bm}
\usepackage{amsmath}

\usepackage[colorlinks,allcolors=blue,linktocpage]{hyperref}
\usepackage{cleveref}
\crefname{equation}{Eq.}{Eqs.}%

\begin{document}
\title{Probing dark gauge boson via Einstein-Cartan portal}
\author{Cao H. Nam}
\email{nam.caohoang@phenikaa-uni.edu.vn}  
\affiliation{Phenikaa Institute for Advanced Study and Faculty of Fundamental Sciences, Phenikaa University, Yen Nghia, Ha Dong, Hanoi 12116, Vietnam}
\date{\today}

\begin{abstract}

Einstein-Cartan gravity which is an alternative formulation of general relativity introduces new degrees of freedom contained in the torsion field which encodes the torsion feature of spacetime. Interestingly, the torsion field couples to all fermions through its axial-vector mode with a universal coupling $\eta=1/8$ which is possible to change under the quantum effects. We argue that Einstein-Cartan gravity provides a significant portal to probe $A'$ dark gauge boson which resides in dark sector existing as an invisible world parallel to our own and couples to the standard model (SM) particles through only the kinetic mixing. For the (very) small kinetic mixing, searches for the $A'$ from Drell-Yan processes are insensitive due to the suppressed production cross-section and the considerable SM backgrounds. However, through the mediation of torsion field the $pp$ collisions produce dark-sector fermions which would significantly produce the $A'$ due to unsuppressed dark gauge coupling. We explore the potential production modes of the $A'$ through bremsstrahlung off dark-sector fermion and the cascade decays. Einstein-Cartan gravity suggests the torsion mass $\gtrsim\mathcal{O}(4)$ TeV for $\eta$ varying around the classical value since the present scenarios would tend to produce the $A'$ with the high boost and large missing transverse momentum from dark-sector fermions where the SM backgrounds are low. On the other hand, the $A'$ search via Einstein-Cartan portal can reach even for the signal events to be not large and is also sensitive to the (very) small kinetic mixing as long as the decay channels of the $A'$ to dark-sector particles are inaccessible.
\end{abstract}

\maketitle

\emph{Introduction.}---The observations in particle physics, astrophysics, and cosmology have motivated the existence of additional $U(1)$ gauge bosons with the (very) small coupling to the standard model (SM) particles \cite{Holdom1986,Galison1984,Fayet1990,GHe1991,Foot1994,EMa2002,Pospelov2008,RChen2009,Arkani-Hamed2009,Hooper2015,Feng2016,Nam2019,Nam2020,Dong2021} (and references therein). In addition, the observations of dark matter (DM) from the gravitational effects provide a strong hint for the presence of dark sector which exists as an invisible world parallel to our own and does not couple directly to the SM particles. This implies that these gauge bosons may reside in dark sector and couple to the SM particles through only the kinetic mixing \cite{Holdom1986,Galison1984}, hence usually referred as dark gauge bosons. The varying range of kinetic mixing parameter $\epsilon$ is rather wide from $\mathcal{O}(10^{-12})$ to $\mathcal{O}(10^{-3})$ realized in various scenarios \cite{Dienes1997,Lukas2000,Abel2004,Goodsell12009}.

Search for dark gauge bosons has been extensively studied at the colliders \cite{CDF2011,CMS2015,ATLAS2017,CMS2020}. However, the direct production cross-section of dark gauge bosons from Drell-Yan processes is suppressed due to the smallness of kinetic mixing parameter and hence the signal is too small to be discovered amid the considerable SM backgrounds, for instance with $\epsilon\lesssim10^{-2}$ \cite{Jaeckel2013,Hoenig2014}. Therefore, there have been proposals for a significant production of dark gauge bosons from dark-sector particles (like DM particles) which are produced from the $pp$ collisions at the LHC \cite{Gupta2015,Autran2015,YBai2015,Buschmann2015,MKim2018,MDu2020}. The final state is the visible decay products of dark gauge bosons into the SM particles plus missing transverse momentum (MET) from dark-sector particles. But, a natural question raised here is what is the portal which allows us to produce dark-sector particles at the colliders, which is essential for the significant production of dark gauge bosons, in the situation that the SM and dark sectors themselves exist in two parallel worlds.\footnote{We have here assumed that the mixing between the dark-sector and SM Higgs bosons is negligibly small because the precise measurements suggest that Higgs boson discovered at the LHC is almost consistent with the SM Higgs boson \cite{Tanabashi2018}.} In this letter, we will point to Einstein-Cartan portal which provides a mechanism of production of dark-sector cascades and represent the corresponding potential avenues for probing dark gauge bosons at hadron colliders.

The choice of the fundamental variables in describing the gravitational dynamics leads to the different formulations of general relativity (GR). In the metric formulation which is the most used approach, the metric field is considered as the fundamental variable and the affine connection $\Gamma^\lambda_{\mu\nu}$ (related to the covariant derivative of tensor fields $\nabla_\mu A^\lambda=\partial A^\lambda+\Gamma^\lambda_{\mu\nu}A^\nu$) is derived from the metric field and its derivatives due to the requirements of $\Gamma^\lambda_{\mu\nu}=\Gamma^\lambda_{\nu\mu}$ and $\nabla_\rho g_{\mu\nu}=0$ (the connection satisfying these conditions is called Christoffel connection or symbol denoted by $\{{\lambda \atop \mu\nu}\}$). As a result, the features of spacetime as well as the gravitational dynamics are manifested by the curvature. An alternative formulation of GR, which is equivalent to the metric formulation in the case of pure gravity or no matter included, is Einstein-Cartan gravity \cite{Kibble1961,Tsoubelis1983,Shaposhnikov2020,Shaposhnikov2021}. In this formulation, the vielbein and the spin connection are the fundamental variables or in other words the metric and affine connection are the independent variables. As a result, the affine connection $\Gamma^\lambda_{\mu\nu}$ possesses an asymmetric part and is decomposed as, $\Gamma^\lambda_{\mu\nu}=\{{\lambda \atop \mu\nu}\}+({{T_\nu}^\lambda}_\mu+{{T_\mu}^\lambda}_\nu+{T^\lambda}_{\mu\nu})/2$, where ${T^\lambda}_{\mu\nu}$ which satisfies ${T^\lambda}_{\mu\nu}=-{T^\lambda}_{\nu\mu}$ is called the torsion field which carries new degrees of freedom characterizing the torsion feature of spacetime. The further explorations and phenomenological implications of torsion field have drawn attention in the literature \cite{Belyaev1998,Belyaev199,Belyaev2007,Skugoreva2015,Boos2017,Almeida2018,Barman2020,Otalora2021,Bahamonde2021,Benisty2021,Dombriz2021}.

\emph{Torsion field and its couplings to matter.}---Torsion field can be decomposed into three irreducible components as follows
\begin{eqnarray}
T_{\lambda\mu\nu}=\frac{1}{3}\left(T_\mu g_{\lambda\nu}-T_\nu g_{\lambda\mu}\right)-\frac{1}{6}\epsilon_{\lambda\mu\nu\rho}S^\rho+q_{\lambda\mu\nu},
\end{eqnarray}
where $T_\mu\equiv{T^\nu}_{\mu\nu}$ is its trace part corresponding to the vector mode, $S^\rho\equiv\epsilon^{\lambda\mu\nu\rho}T_{\lambda\mu\nu}$ is its completely antisymmetric part corresponding to the axial-vector mode, and $q_{\lambda\mu\nu}$ is its tracefree part satisfying the conditions ${q^\nu}_{\mu\nu}=0$ and $\epsilon^{\mu\nu\rho\lambda}q_{\mu\nu\rho}=0$.

The spin connection does not vanish in the Minkowski-flat spacetime ($g_{\mu\nu}=\eta_{\mu\nu}$) with the presence of torsion field. As a result, it leads to the couplings of some irreducible component of torsion field to any Dirac fermion with a general Lagrangian read
\begin{eqnarray}
\mathcal{L}_D=\bar{\psi}i\gamma^\mu\left(\partial_\mu+i\eta\gamma^5S_\mu+\cdots\right)\psi-m\bar{\psi}\psi,
\end{eqnarray}
where $\eta=1/8$ is the prediction of Einstein-Cartan gravity and the ellipse refers to the terms relating to the gauge fields of the SM gauge symmetry. We see here that the torsion field couples to the fermions which are both SM and hidden fermions through its axial-vector part with a universal coupling $\eta=1/8$ at the classical level. This suggests that although the SM and dark sectors themselves exist in two parallel worlds, the universal coupling of torsion field to any fermion is crucial for a production of dark-sector fermions, from the collisions of the SM fermions, which subsequently radiate or decay into lighter dark-sector particles to significantly produce dark gauge bosons with an unsuppressed production cross-section. 

In the situation of dynamical torsion field, the classical value $\eta=1/8$ is no longer stable under the quantum corrections and $\eta$ runs in terms of the energy scale. Therefore, we consider the fermion-torsion coupling $\eta$ as a free parameter in the following analysis.

One can establish the action for the dynamical torsion field in the context of effective low-energy quantum theory using the consistency criteria of unitarity and renormalizability \cite{Donoghue1992,Weinberg1995}. As seen above, the trace and tracefree parts of torsion field decouple to theory, hence we set them to be zero and consider only its axial-vector part to be present. This means that the torsion field can be parameterized as $T_{\lambda\mu\nu}=-\epsilon_{\lambda\mu\nu\rho}S^\rho/6$. In this way, we here consider a sub-class of most general torsion fields where the torsion field is totally antisymmetric in all three indices. Since, in the framework of effective field theory the action of the dynamical torsion field would contain the terms related to the derivatives in $S_\mu$. In this way, the action of Einstein-Cartan gravity which is obtained by extending the Einstein-Hilbert action with including the dynamical torsion field is given by
\begin{eqnarray}
S&=&\frac{M^2_{\text{Pl}}}{2}\int d^4x\sqrt{-g}\left[\mathcal{R}(\Gamma)+\frac{1}{\Lambda^2}\left(c_1S_{\mu\nu}S^{\mu\nu}\right.\right.\nonumber\\
&&\left.\left.+c_2(\partial_\mu S^\mu)^2+\text{higher-order terms}\right)\right],\label{extEHact}
\end{eqnarray}
where $\mathcal{R}(\Gamma)$ is the Ricci scalar written in terms of the affine connection $\Gamma$, $M_{\text{Pl}}$ is the Planck scale, $S_{\mu\nu}\equiv\partial_\mu S_\nu-\partial_\nu S_\mu$, $\Lambda$ is an energy scale below which the effects of torsion field are negligible, and $c_{1,2}$ are the parameters. Note that, the higher-order terms here are highly suppressed by the Planck mass. We can split the Ricci scalar $\mathcal{R}(\Gamma)$ into two parts as $\mathcal{R}(\Gamma)=\bar{\mathcal{R}}+S_\mu S^\mu/4$ where $\bar{\mathcal{R}}$ is the usual Ricci scalar corresponding to the Christoffel connection and the second term is related to the torsion mass. Whereas, in order to have an effective low-energy quantum theory of torsion field compatible with the requirements of unitarity and renormalizability, the parameter $c_1$ must be nonzero and negative which is suitably chosen to be $-1/4$ and the parameter $c_2$ must be zero \cite{Belyaev1998,Belyaev199}. By rescaling the torsion field as $S_\mu\rightarrow\sqrt{2}\Lambda S_\mu/M_{\text{Pl}}$, we find Lagrangian of torsion field in terms of its axial-vector part as 
\begin{eqnarray}
\mathcal{L}_T=-\frac{1}{4}S_{\mu\nu}S^{\mu\nu}+\frac{1}{2}m^2_TS_\mu S^\mu,
\end{eqnarray}
where the torsion mass is determined by $m_T=\Lambda/\sqrt{2}$.

Besides the metric field corresponding to Einstein gravity, the low-energy limit of string theory also predicts an antisymmetric torsion field related to Kalb-Ramond field. Thus, within string-inspired Einstein-Cartan gravity the torsion mass is expected to be of the Planck mass order. Such a torsion scenario was considered in \cite{Mavromatos2012} where the authors indicated that the gravitational interactions related to the torsion field allow to generate the right-handed Majorana neutrino masses at two loops involving the global anomalies. However, considering the torsion mass of the Planck mass order would be less appealing for the collider phenomenology because the torsion interactions are suppressed by the Planck scale and hence the effects of torsion field are practically unobservable. In the present work, we suppose that the torsion field propagates at the energies which are much lower than the Planck scale and available at the present/future colliders.

\emph{Simplified $U(1)_D$ model with Einstein-Cartan portal.}---In the presence of torsion field, we extend the SM with dark sector which is represented by $A'_\mu$ dark gauge boson corresponding to an extra $U(1)_D$ gauge group and a Dirac fermion $\chi$. The corresponding Lagrangian is given by
\begin{eqnarray}
\mathcal{L}=\mathcal{L}_D+\mathcal{L}_T+\mathcal{L}_{\text{gauge}}+\bar{\chi}\left(i\gamma^\mu D_\mu-m_\chi\right)\chi,
\end{eqnarray} 
where $D_\mu=\partial_\mu+i\eta\gamma^5S_\mu+ig_DA'_\mu$ with $g_D$ to be the dark gauge coupling corresponding to $U(1)_D$ and $\mathcal{L}_{\text{gauge}}$ reads
\begin{eqnarray}
\mathcal{L}_{\text{gauge}}&=&-\frac{1}{4}B_{\mu\nu}B^{\mu\nu}-\frac{1}{4}F'_{\mu\nu}F'^{\mu\nu}+\frac{\epsilon}{2c_W}F'_{\mu\nu}B^{\mu\nu}\nonumber\\
&&+\frac{m^2_{A'}}{2}A'_\mu A'^\mu,
\end{eqnarray}
here $B_{\mu\nu}$ and $F'_{\mu\nu}$ is the field strength tensors corresponding to the $B_\mu$ and $A'_\mu$ gauge fields of the $U(1)_Y$ and $U(1)_D$ gauge groups, respectively, $c_W$ is the cosine of Weinberg angle, and $m_{A'}$ is the $A'$ mass which can be generated from the nonzero vacuum expectation value (VEV) of a dark-sector scalar field. Well motivated by the theoretical and observational reasons which dark sector may exist around the electroweak scale, we are here concerned in the mass range of the $A'$ which is from a few tens of GeV to a few hundreds of GeV. This mass range should correspond to the promptly decaying or short-lived $A'$ if the kinetic mixing parameter is small enough. 

By canonically normalizing the gauge fields and rotating the mass eigenstates, one obtains the coupling of the $A'$ to the SM fermions as \cite{He2018}
\begin{eqnarray}
\mathcal{L}\supset\epsilon e\left[Q_f\bar{f}\gamma^\mu f+\frac{\hat{g}}{2c^2_W}\bar{f}\gamma^\mu(g^f_V-g^f_A\gamma^5)f\right]A'_\mu,
\end{eqnarray}
where $Q_f$ is the electric charge of fermion $f$, $\hat{g}\equiv m^2_{A'}/(m^2_Z-m^2_{A'})$, $g^f_V=T^f_3-2s^2_WQ_f$, and $g^f_A=T^f_3$ with $T^f_3$ to be the weak isospin of fermion $f$. The kinetic mixing leads also to the additional couplings of $\chi$ to the $Z$ boson of the SM as $\mathcal{L}\supset-(g_Dt_W\epsilon\hat{g}m^2_Z/m^2_{A'})\bar{\chi}\gamma^\mu\chi Z_\mu$.

We discuss the constraint on the torsion mass from the LHC limits \cite{ATLAS2017}. For the fermion-torsion coupling which varies around the classical value predicted by Einstein-Cartan gravity, one can find a lower bound of $m_T$ which is $\mathcal{O}(4)$ TeV for the present model. However, the bound can relaxed if we consider many additional fermions with their masses smaller than $m_T/2$, which would reduce the branching ratio $\text{Br}(TS\rightarrow l^+l^-)$.

\emph{Relic abundance and direct detection.}---The pair-annihilation of $\chi$ into the SM fermion pairs is mediated through the $s$-channel exchange of the $A'$, $Z$, and torsion field. However, these annihilation channels are sub-dominant because they are strongly suppressed by the small kinetic mixing and $m^2_f/m^2_T$ where $m_f$ refers to the mass of the SM fermions. Therefore, the dominant pair-annihilation of $\chi$ which contributes to its relic abundance is into the $A'$ pair via the $t$-channel. The thermally averaged cross-section of this pair-annihilation is given as $\langle\sigma v\rangle=g^4_D(1-x^2)^{3/2}(1-x^2/2)^{-2}/(16\pi m^2_\chi)$ where $x\equiv m_{A'}/m_\chi$. The relic abundance of $\chi$ is computed as $\Omega_{\chi}h^2\simeq2.12\times10^{-10}\text{GeV}^{-2}/\langle\sigma v\rangle$. As seen below, in order to obtain the sufficiently large $A'$ production cross-section through bremsstrahlung, the dark gauge coupling is as sizable as $\mathcal{O}(1)$. Since the $\langle\sigma v\rangle$ annihilation cross-section is very large and the relic abundance of $\chi$ is very small compared to the observed DM abundance $\Omega_{\text{DM}}h^2\simeq0.12$ \cite{Planck2016}. Accordingly, the contribution of $\chi$ to DM almost agrees with the cosmological constraints on the fraction of millicharged DM which is less than $1\%$ for $m_\chi<100$ GeV \cite{Boddy2018,Munoz2018,dePutter2019}.

The $\chi$-nucleon scattering occurs through the $t$-channel exchange of the $A'$, $Z$, and torsion field. The contribution of the $A'$ and $Z$ gives rise the spin-independence (SI) scattering because the interactions of $\chi$ to the $A'$ and $Z$ are described by the vectorial-vector couplings. Whereas, the axial-vector couplings of torsion field to both $\chi$ and SM charged fermions will lead to the spin-dependence (SD) scattering. If $\chi$ constitutes all of the DM, it would be constrained by the direct-detection experiments. The constraints on SI scattering cross-section are given by several experiments \cite{PandaXII,XENON1T}, which essentially impose the bounds on the kinetic mixing parameter. A detail study can be found in Ref. \cite{Chun2011}. The most stringent constraint on SD scattering cross-section comes from the LUX experiment \cite{LUX2016}. On the contrary, if $\chi$ makes up only a small fraction of the DM abundance, these constraints can be relaxed.

\begin{figure}[t]
 \centering
\begin{tabular}{cc}
\includegraphics[width=0.45 \textwidth]{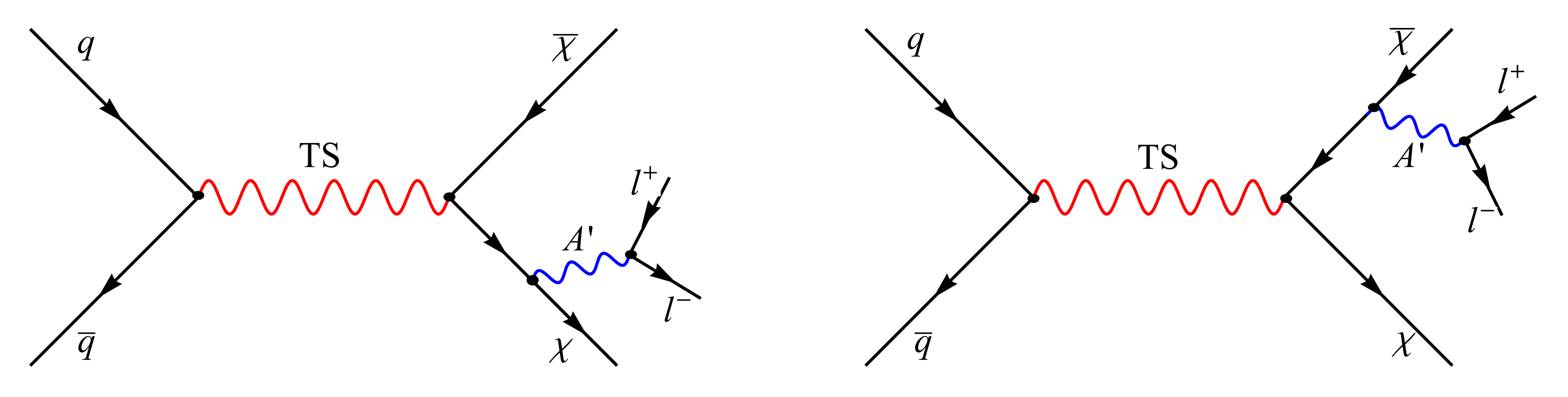}
\end{tabular}
 \caption{The $A'$ production in the final state radiation and its subsequent decay into the lepton pair where $\text{TS}$ stands for the torsion field.}\label{DGB-pro}
\end{figure}

\emph{The $A'$ production through bremsstrahlung.}---Einstein-Cartan portal provides a potential production mode of the $A'$ through bremsstrahlung off dark-sector fermion $\chi$ whose production in the $pp$ collisions is via the mediation of torsion field. The $A'$ subsequently decays to the SM fermion pairs with the significant branching ratios if $m_{A'}<2m_\chi$, where the dilepton decays would be the primary channels for the $A'$ search. This process would lead to an observable dilepton resonance in association with MET from dark-sector fermions. The Feynman diagram of this process is shown in Fig. \ref{DGB-pro}.

The corresponding cross-section for on-shell torsion field can be written as
\begin{eqnarray}
\sigma&=&\sigma(pp\rightarrow TS\rightarrow l^+l^-)R_{A'}\text{Br}(A'\rightarrow l^+l^-),\label{darkstr-sigma}
\end{eqnarray}
where we have considered the decay channels of $e^+e^-$ and $\mu^+\mu^-$ combined, $\sigma(pp\rightarrow TS\rightarrow l^+l^-)$ is the cross-section of process $pp\rightarrow TS\rightarrow l^+l^-$ whose upper bound is about $0.2$ fb for $m_T\gtrsim3$ TeV \cite{ATLAS2017}, the $R_{A'}$ is defined as
\begin{eqnarray}
R_{A'}\equiv\frac{\sigma(pp\rightarrow TS\rightarrow\bar{\chi}\chi A')}{\sigma(pp\rightarrow TS\rightarrow l^+l^-)}=\frac{\Gamma(TS\rightarrow\bar{\chi}\chi A')}{\Gamma(TS\rightarrow l^+l^-)},
\end{eqnarray}
which depends on $g_D$, $m_T$, $m_{A'}$, and $m_\chi$, and $\text{Br}(A'\rightarrow l^+l^-)$ is the branching ratio of the $A'$ decay to $e^+e^-$ and $\mu^+\mu^-$. We see that, if the decay channels of the $A'$ to dark-sector particles are inaccessible, the cross-section of the above process depends on the $A'$ and torsion masses, the fermion-torsion coupling, the dark gauge coupling, and the branching ratio $\text{Br}(A'\rightarrow l^+l^-)$, without depending on the kinetic mixing parameter.

\begin{figure}[t]
 \centering
\begin{tabular}{cc}
\includegraphics[width=0.4 \textwidth]{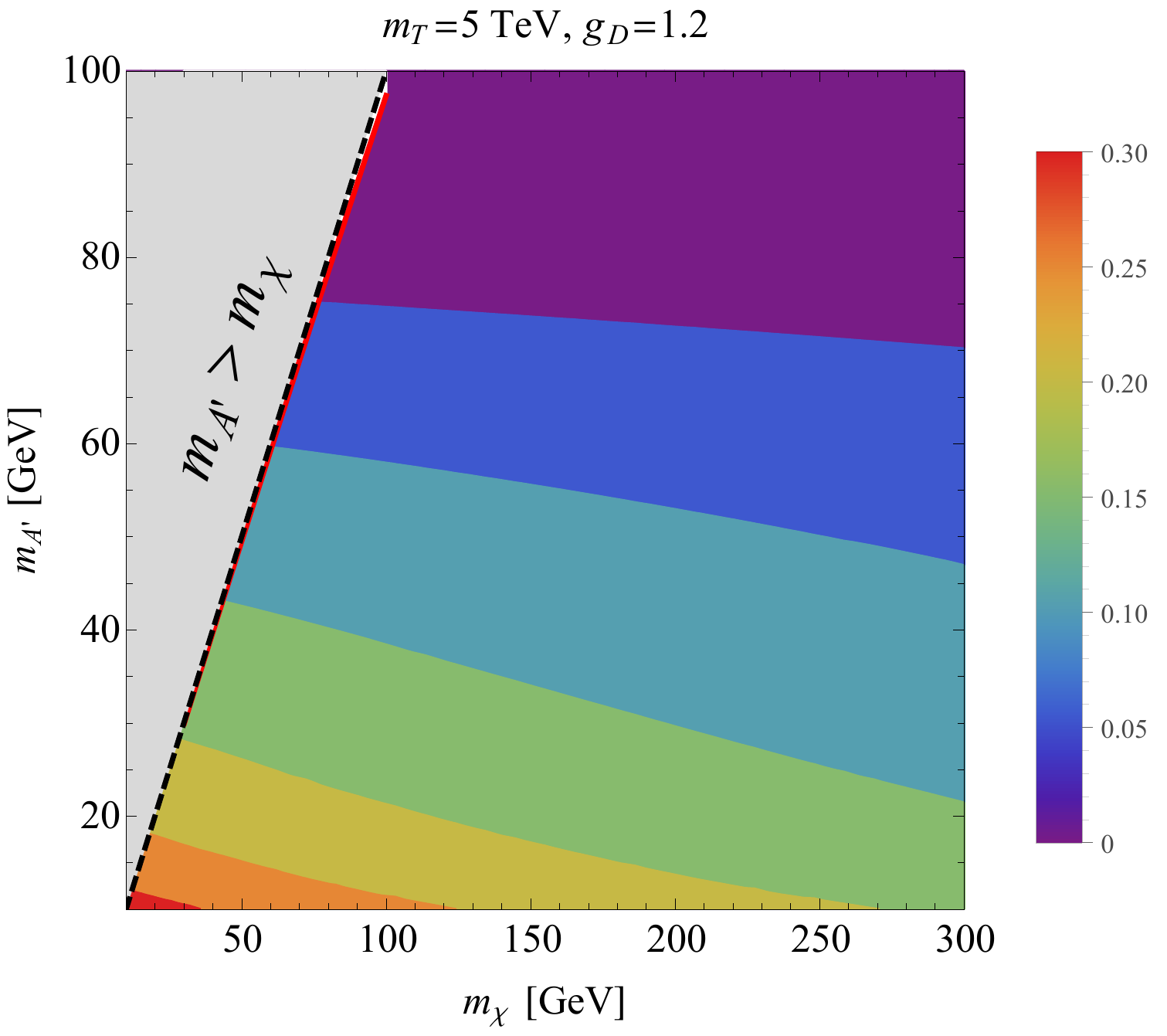}
\end{tabular}
 \caption{The contour of product $R_{A'}\text{Br}(A'\rightarrow l^+l^-)$. The red line refers to an upper bound corresponding to the cosmological constraints on the fraction of millicharged DM for $m_\chi<100$ GeV \cite{Boddy2018,Munoz2018,dePutter2019}.}\label{RAtimesBr}
\end{figure}

The signal of the above process is very similar to that coming from the decay of charginos pair into the lepton pair plus MET. The null results from the search for electroweakinos carried out by ATLAS \cite{ATLAS2020} can be used to place an upper bound on the cross-section $\sigma(pp\rightarrow TS\rightarrow\bar{\chi}\chi A')$ times branching ratio $\text{Br}(A'\rightarrow l^+l^-)$ which is at the level of $\mathcal{O}(0.1)$ fb. As mentioned above, for the fermion-torsion coupling varying around the value predicted by Einstein-Cartan gravity, the LHC constraint requires $m_T\gtrsim\mathcal{O}(4)$ TeV corresponding to $\sigma(pp\rightarrow TS\rightarrow l^+l^-)\lesssim 0.2$ fb. This along with $R_{A'}\text{Br}(A'\rightarrow l^+l^-)\lesssim0.25$ for the short-lived $A'$ under consideration (see Fig. \ref{RAtimesBr}) points to that $\sigma(pp\rightarrow TS\rightarrow\bar{\chi}\chi A')$ times $\text{Br}(A'\rightarrow l^+l^-)$ is less $0.05$ fb and thus it is well consistent with the existing LHC constraint.

The $A'$ production at the LHC via the mediation of torsion field which belongs to the TeV-scale mass region tends to produce the $A'$ with the high boost and large MET where the number of the background events is low and hence it increases the signal sensitivity of the $A'$ search in this process.\footnote{The main backgrounds at the LHC for the dilepton final state plus MET are $pp\rightarrow W^+W^-$, $W^{\pm}Z$, $ZZ$, and $t\bar{t}$ \cite{ATLAS2020}.} This suggests that applying the high selection cuts on the transverse momentum $p_{T,ll}$ of the lepton pair and MET can almost completely eliminate the background events but it does not lead to a significant reduction in the signal efficiency. On the other hand, the $A'$ search can reach even for the number of signal events to be not large. Furthermore, the contour plot of $R_{A'}\text{Br}(A'\rightarrow l^+l^-)$ in the $m_\chi-m_{A'}$ plane given in Fig. \ref{RAtimesBr} implies that the better signal sensitivity can reach with the sufficiently high integrated luminosities which are expected to achieve at the future LHC.

In order to see the potential parameter region of the present scenario at the $14$ TeV LHC with the highly integrated luminosity, we show the projections in the $m_\chi-m_{A'}$ plane with requiring benchmark number of signal events $N=10$ as the criterion of the LHC reach. Here, the expected number of signal events is computed as $N=\epsilon\mathcal{L}\sigma$ with $\sigma$ given in Eq. (\ref{darkstr-sigma}), $\mathcal{L}$ to be integrated luminosity, and $\epsilon$ denoting the signal efficiency. As discussed, the signal in the present scenario is highly efficient even at the high selection cuts. Hence, we can consider a constant signal efficiency $\epsilon\approx100\%$.
\begin{figure}[t]
 \centering
\begin{tabular}{cc}
\includegraphics[width=0.4 \textwidth]{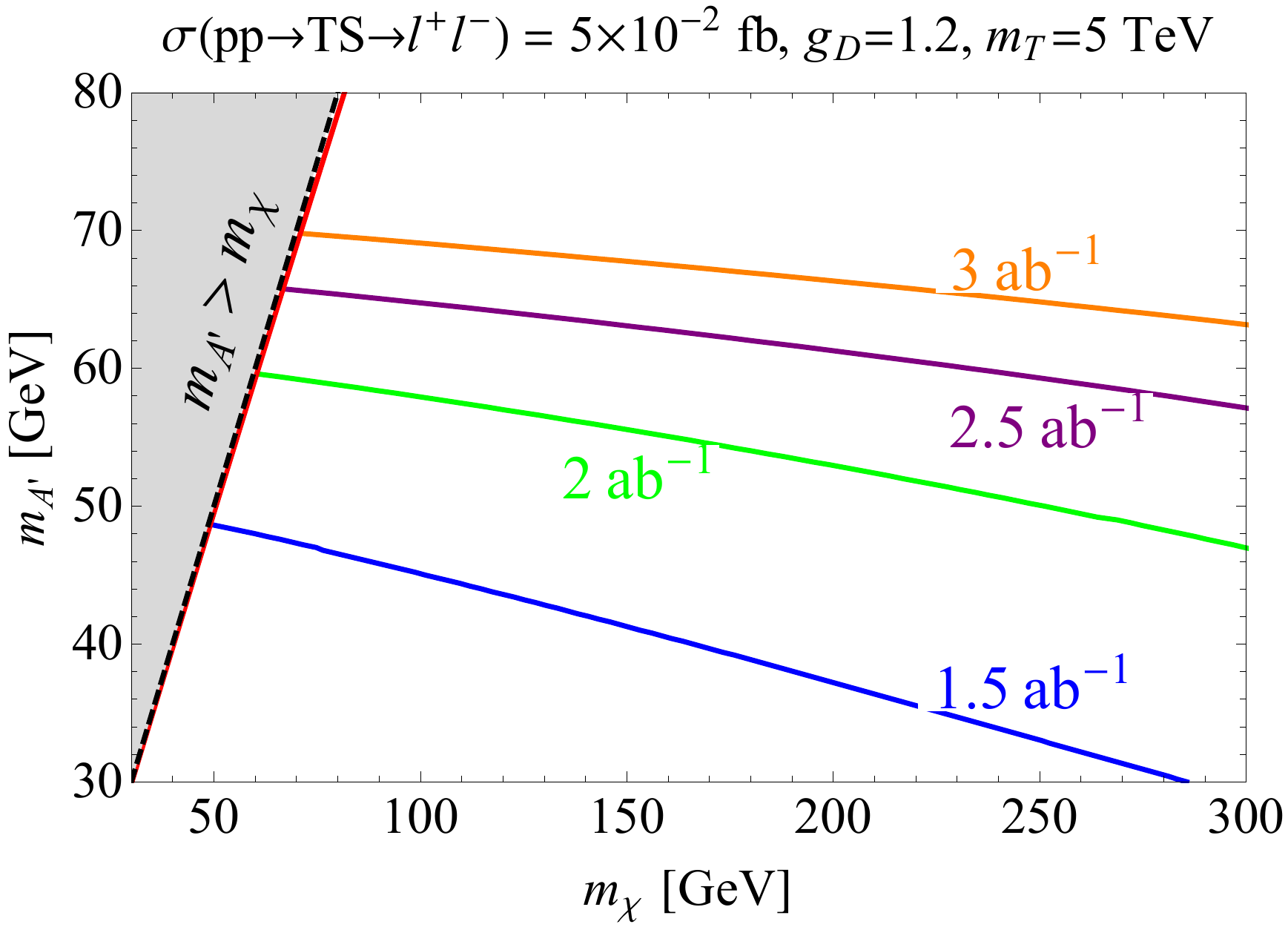}
\end{tabular}
 \caption{The contour of signal events for the $14$ TeV LHC reach for various values of integrated luminosity at the future LHC. The meaning of the red line is the same as in Fig. \ref{RAtimesBr}.}\label{Darkstrahlung}
\end{figure}
We find that for the integrated luminosity which is about one-half of that at the HL-LHC, it will be possible to probe the low mass region of the short-lived $A'$. Its higher mass region can be probed with increasing the integrated luminosity, for example, at the HL-LHC the short-lived $A'$ in the mass region around $70$ GeV can be enhanced over the SM backgrounds.

\emph{The $A'$ production through cascade decays.}---We discuss another potential avenue for probing the $A'$: through the mediation of torsion field the $pp$ collisions produce the heavier dark-sector fermion pairs each of which would decay into a lighter dark-sector fermion and the $A'$ which decays promptly to the SM fermion pairs. We consider the situation that the left- and right-handed components of Dirac fermion $\chi$ have the Yukawa couplings to a dark-sector scalar field $\phi$ as $-(y_1\bar{\chi}_L\chi^c_L\phi+y_2\bar{\chi}_R\chi^c_R\phi+\text{h.c.})/2$. When $\phi$ develops VEV, $\chi_{L,R}$ acquire the Majorana masses besides the mass of Dirac type. As a result, the mass diagonalization splits the original Dirac fermion $\chi$ into two Majorana fermions denoted by $\chi_1$ and $\chi_2$ whose masses are given by $m_1\simeq m_D-m_+$ and $m_2\simeq m_D+m_+$, respectively, where $m_\pm\equiv m_L\pm m_R$ with $m_{L,R}=h_{1,2}\langle\phi\rangle$ and we have assumed $m_-/m_D\ll1$. The couplings of torsion field and the $A'$ to two mass eigenstates $\chi_{1,2}$ are found as
\begin{eqnarray}
\mathcal{L}_{\text{int}}&\supset&-\eta\left(\bar{\chi}_1\gamma^\mu\chi_1+\bar{\chi}_2\gamma^\mu\chi_2\right)S_\mu-g_D\left[\bar{\chi}_1\gamma^\mu\chi_2+\text{h.c.}\right.\nonumber\\
&&\left.+\frac{m_-}{m_D}\left(\bar{\chi}_2\gamma^\mu\chi_2-\bar{\chi}_1\gamma^\mu\chi_1\right)\right]A'_\mu.\label{TS-Ap-DM}
\end{eqnarray}
We observe here that the interaction of torsion field to two mass eigenstates $\chi_{1,2}$ only is defined by the diagonal couplings, whereas, the dominant interaction of the $A'$ to $\chi_{1,2}$ is off-diagonal and the diagonal couplings are suppressed by $m_-/m_D$. The difference here is due to the fact that the coupling forms of torsion field and the $A'$ to the original Dirac fermion $\chi$ are axial-vector and vectorial-vector, respectively.

\begin{figure}[t]
 \centering
\begin{tabular}{cc}
\includegraphics[width=0.35 \textwidth]{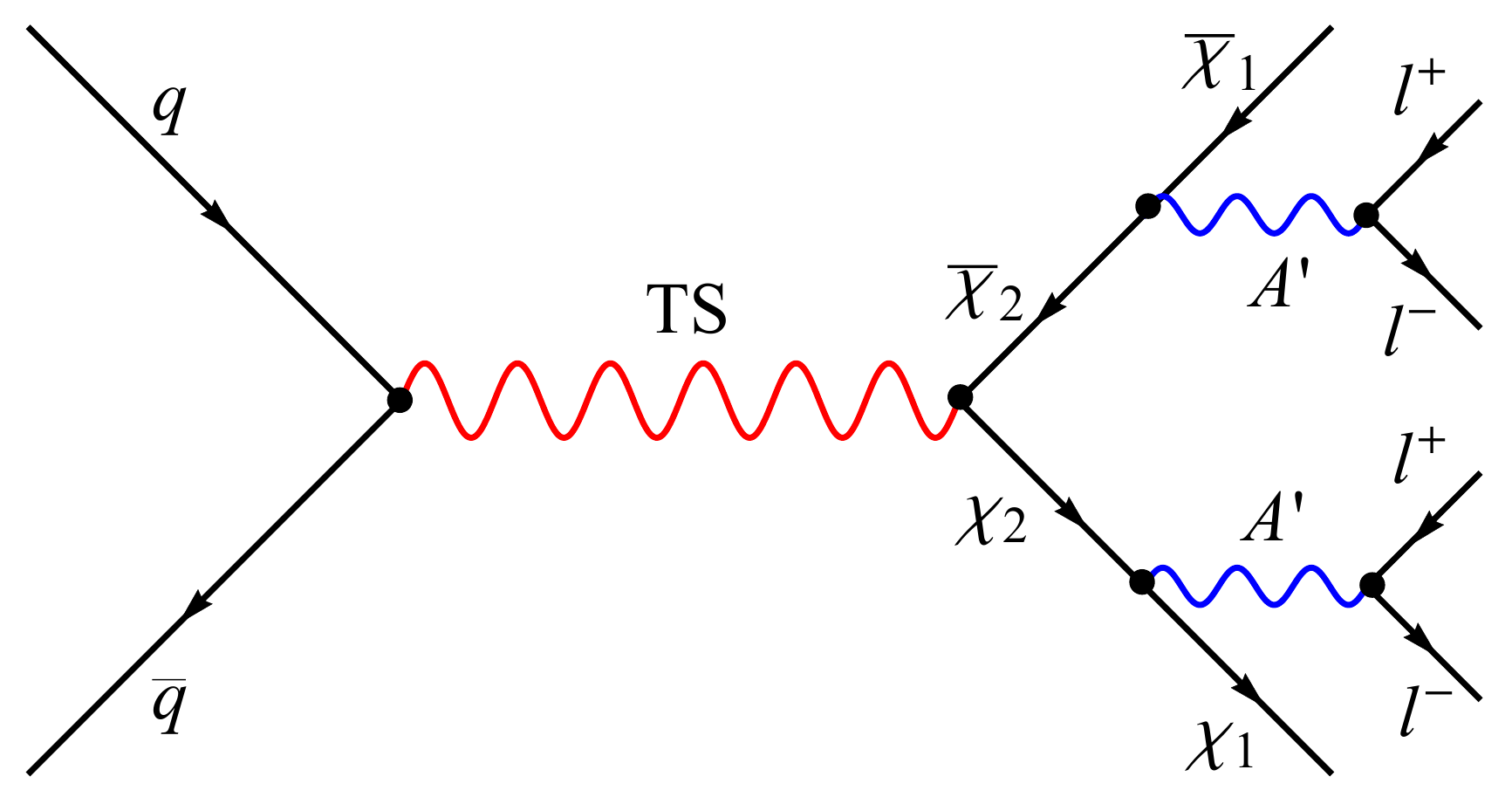}
\end{tabular}
 \caption{The $A'$ production from the cascade decay of $\chi_2$ heavier dark-sector fermion into another $\chi_1$ lighter one and the $A'$ decays into the dilepton.}\label{CDC-pro}
\end{figure}

From the coupling Lagrangian (\ref{TS-Ap-DM}), one can realize another production mode of the $A'$ in the $pp$ collisions. Through the Drell-Yan process mediated by the torsion field, $\chi_2$ is pair produced from the $pp$ collisions each of which would decay into the $\chi_1$ and the $A'$. The $A'$ decays promptly into the SM fermion pairs where we are interested in the decay channels of $e^+e^-$ and $\mu^+\mu^-$ combined. This process would be in association with MET and four leptons in the final state and is illustrated in Fig. \ref{CDC-pro}. 
The corresponding cross-section for on-shell torsion field is written as
\begin{eqnarray}
\sigma&\simeq&\frac{1}{4}\sigma(pp\rightarrow TS\rightarrow l^+l^-)\left[\text{Br}(\chi_2\rightarrow\chi_1 A')\right]^2\nonumber\\
&&\times\left[\text{Br}(A'\rightarrow l^+l^-)\right]^2,\label{prcs-casdec}
\end{eqnarray}
which depends on the fermion-torsion coupling, the torsion mass, and the branching ratios of $TS\rightarrow\bar{\chi}_2\chi_2$, $\chi_2\rightarrow\chi_1A'$, and $A'\rightarrow l^+l^-$. We have here assumed that the masses of the SM and dark-sector fermions are much smaller than the torsion mass and thus they are negligible in the decay widths of torsion field. 

In particular, when the $\chi_2$ decays into only the $\chi_1$ and the $A'$, i.e. $\text{Br}(\chi_2\rightarrow\chi_1 A')=1$, the cross-section given in Eq. (\ref{prcs-casdec}) is independent on the dark gauge coupling and the masses of the $\chi_{1,2}$. Consequently, we can identify the $\chi_1$ as DM candidate because its relic abundance may make up all of DM. This is an advantage of the $A'$ production scenario through the cascade decays compared to through bremsstrahlung where the relic abundance of $\chi$ is too small due to the $\mathcal{O}(1)$ sizable dark gauge coupling. 

In analogy to the $A'$ production mode through bremsstrahlung discussed previously, the $A'$ production through the cascade decays resulting in the four-lepton final state plus MET would tend to produce the $A'$ with the high boost and large MET. Thus, applying the high selection cuts can drastically suppress the backgrounds without significantly reducing in the signal efficiency.

\begin{figure}[t]
 \centering
\begin{tabular}{cc}
\includegraphics[width=0.4 \textwidth]{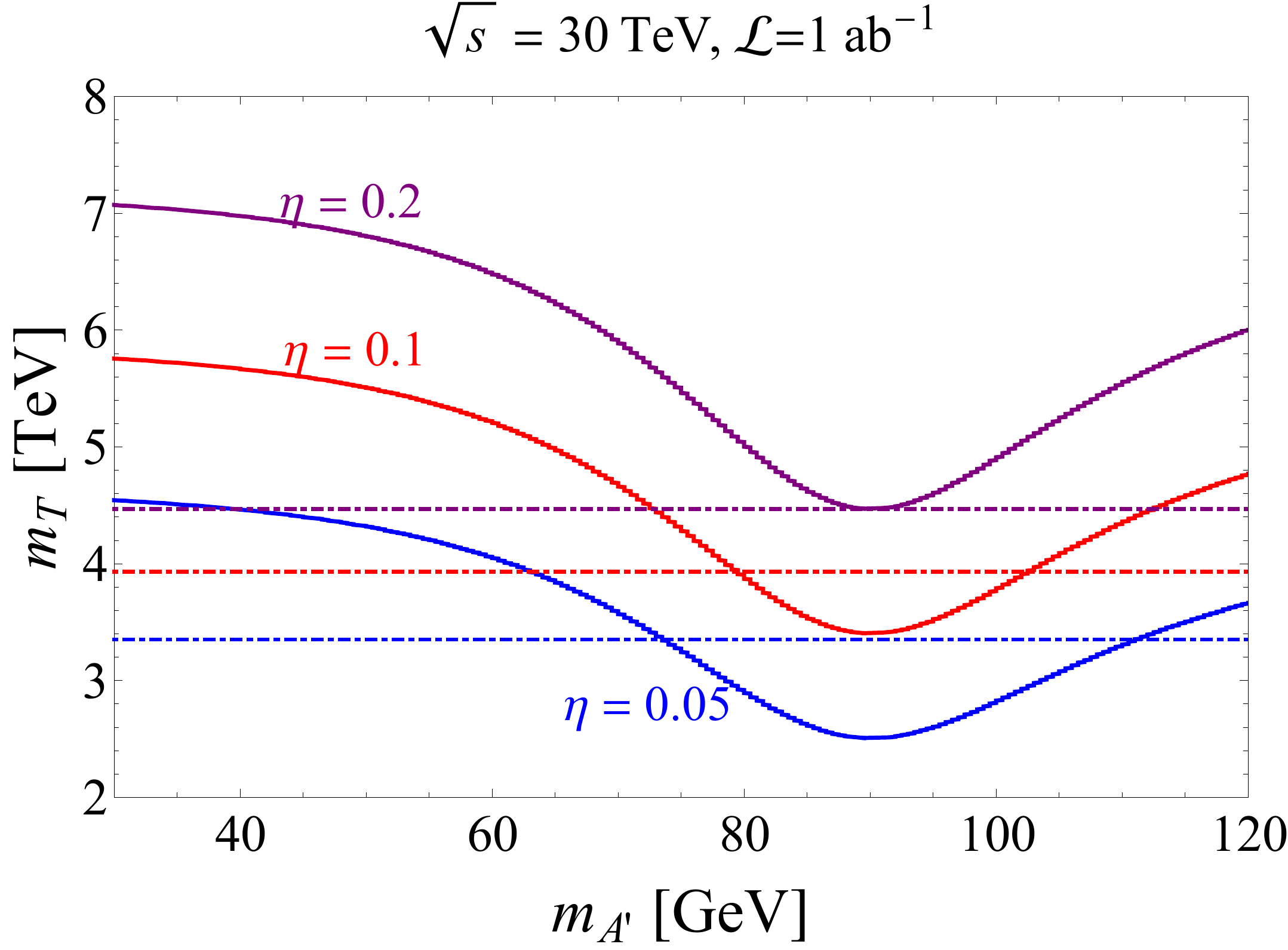}
\end{tabular}
 \caption{The contour of signal events for the reach of future collider with $\sqrt{s}=30$ TeV and $\mathcal{L}=1$ ab$^{-1}$ for various values of fermion-torsion coupling. The blue, red, and purple dashed lines refer to the existing LHC lower bounds corresponding to $\eta=0.05$, $0.1$, and $0.2$, respectively.}\label{Cascade-decays}
\end{figure}

For $\sigma(pp\rightarrow TS\rightarrow l^+l^-)$ close to the existing LHC upper bound $\lesssim0.2$ fb for $m_T\gtrsim3$ TeV and from $\text{Br}(A'\rightarrow l^+l^-)\lesssim0.3$, we find that the signal is observable at the integrated luminosities around the value achieved at the HL-LHC and in the low $A'$ mass region. On the other hand, the signal in the $A'$ production through the cascade decays is relatively less sensitive at the 14 TeV LHC in a wide range of parameters. Therefore, we study the signal sensitivity of this $A'$ production mode at a future collider with the colliding energy above 14 TeV. In Fig. \ref{Cascade-decays}, we show the potential parameter region which will be able to be probed at the future collider at $\sqrt{s}=30$ TeV and $\mathcal{L}=1$ ab$^{-1}$ with requiring benchmark number of signal events $N=10$ as the reach criterion.
We find that for the sufficiently heavy torsion field we will be able to probe almost the mass region of the short-lived $A'$. On the contrary, the sensitivity can visibly reach for its low and high mass regions. This is due to the fact that the branching ratio $\text{Br}(A'\rightarrow l^+l^-)$ is small in the middle region around $90$ GeV.

\emph{Discussion.}---In this letter, we have pointed to that Einstein-Cartan gravity provides a portal for a significant production of $A'$ dark gauge boson which corresponds to a $U(1)_D$ symmetry in dark sector and couples to the SM particles through only the kinetic mixing. Einstein-Cartan gravity introduces new degrees of freedom contained in the torsion field which couples to all fermions through its axial-vector mode with a universal coupling which is predicted to be $1/8$ in Einstein-Cartan gravity but can vary under the quantum corrections. Unlike the direct $A'$ production from Drell-Yan processes which is strongly suppressed by the (very) small kinetic mixing, the $A'$ is significantly produced in the processes that dark-sector fermions are produced from the $pp$ collisions through the exchange of torsion field and then radiate the $A'$ or decay to lighter dark-sector fermions and the $A'$ if the $A'$ decays to only the SM fermion pairs. Due to the torsion mass in the TeV-scale regime for $\eta$ varying around the classical value, the $A'$ is produced with the high boost and MET from dark-sector fermions is large where the SM backgrounds are low. Hence the search for the $A'$ can reach even for the signal events to be not large and is also sensitive to the (very) small kinetic mixing.

We can extend this work for the light $A'$ which has the mass in the range from MeV to a few GeV. With such masses and the mediation of TeV-scale torsion field, the $A'$ produced from the above discussed scenarios can be very highly boosted and hence it would travel for some macroscopic distance larger than $\mathcal{O}(1)$ mm before decaying. As a result, the $A'$ decay point is significantly displaced from its production point. The corresponding decays to the $l^+l^-$ pairs ($l=e$ or $\mu$) and the hadronic resonances are detected by the timing detectors which can be used to search for the long-lived particles \cite{Liu2019,Cerri2019,Mason2019,Flowers2020} and have been recently proposed to install at the LHC \cite{CMS2017,ATLAS2018,LHCb2017}. Since the probability of emitting the $A'$ (or the $R_{A'}$) and the branching ratios of the $A'$ into the leptonic and hadronic final states are rather large for the light $A'$, the signal event number would be substantially enhanced.

Einstein-Cartan gravity which is here studied can also provide a portal to search for the light $A'$ in the exotic signature of the direct-detection experiments of DM \cite{DKim2019}. In the scattering of cosmogenic boosted DM off nucleus or electron via the $t$-channel exchange of torsion field, the scattered/incident boosted DM may radiate the $A'$. The resultant final state would be a target recoil along with visible decay products of the emitted $A'$ into the SM particles. The cross-section of this process is larger with decreasing the $A'$ and boosted DM masses and increasing the dark gauge coupling and the incoming energy of boosted DM. This suggests that the sufficiently large incoming energy of boosted DM and the sizable dark gauge coupling may lead to a significant event rate between this signal process and the conventional elastic scattering of boosted DM.

We thank Dr. Van Que Tran (Nanjing University) for his talk and discussions about Ref. \cite{MDu2020}. This research is funded by Vietnam National Foundation for Science and Technology Development (NAFOSTED) under grant number 103.01-2019.353.

\end{document}